\documentclass[amssymb,amsmath,aps,prl,twocolumn,superscriptaddress,showpacs]{revtex4}

\usepackage{graphicx}
\usepackage{dcolumn}
\usepackage{bm}

\begin{document}

\title{Evolution of magnetic phases and orbital occupation in (SrMnO$_{3}$)$_{n}$/(LaMnO$_{3}$)$_{2n}$ superlattices}

\author{C. Aruta}
\email{aruta@na.infn.it} \affiliation{CNR-INFM Coherentia,
Complesso Universitario di Monte Sant'Angelo, Via Cintia, I-80125
Napoli, Italy}
\author{C. Adamo}
\affiliation{Cornell University, Ithaca, New York 14853-1501,
U.S.A.}
\author{A. Galdi}
\affiliation{CNR-INFM Coherentia, Complesso Universitario di Monte
Sant'Angelo, Via Cintia, I-80125 Napoli, Italy}
\affiliation{Dipartimento di Scienze Fisiche, Universit\`{a} di
Salerno, Baronissi (SA), Italy}
\author{P. Orgiani}
\affiliation{CNR-INFM Coherentia, Complesso Universitario di Monte
Sant'Angelo, Via Cintia, I-80125 Napoli, Italy}
\affiliation{Dipartimento di Matematica ed Informatica,
Universit\`{a} di Salerno, I-84081 Baronissi (SA), Italy}
\author{V. Bisogni}
\affiliation{European Synchrotron Radiation Facility, Bo\^{\i}te
Postale 220, F-38043 Grenoble, France}
\author{N. B. Brookes}
\affiliation{European Synchrotron Radiation Facility, Bo\^{\i}te
Postale 220, F-38043 Grenoble, France}
\author{J. C. Cezar}
\affiliation{European Synchrotron Radiation Facility, Bo\^{\i}te
Postale 220, F-38043 Grenoble, France}
\author{P. Thakur}
\affiliation{European Synchrotron Radiation Facility, Bo\^{\i}te
Postale 220, F-38043 Grenoble, France}
\author{C. A. Perroni}
\affiliation{CNR-INFM Coherentia, Complesso Universitario di Monte
Sant'Angelo, Via Cintia, I-80125 Napoli, Italy}
\affiliation{Universit\`{a} di Napoli Federico II, Complesso
Universitario di Monte Sant'Angelo, Via Cintia, Napoli, Italy}

\author{G. De Filippis}
\affiliation{CNR-INFM Coherentia, Complesso Universitario di Monte
Sant'Angelo, Via Cintia, I-80125 Napoli, Italy}
\affiliation{Universit\`{a} di Napoli Federico II, Complesso
Universitario di Monte Sant'Angelo, Via Cintia, Napoli, Italy}

\author{V. Cataudella}
\affiliation{CNR-INFM Coherentia, Complesso Universitario di Monte
Sant'Angelo, Via Cintia, I-80125 Napoli, Italy}
\affiliation{Universit\`{a} di Napoli Federico II, Complesso
Universitario di Monte Sant'Angelo, Via Cintia, Napoli, Italy}

\author{D. G. Schlom}
\affiliation{Cornell University, Ithaca, New York 14853-1501,
U.S.A.}
\author{L. Maritato}
\affiliation{CNR-INFM Coherentia, Complesso Universitario di Monte
Sant'Angelo, Via Cintia, I-80125 Napoli, Italy}
\affiliation{Dipartimento di Matematica ed Informatica,
Universit\`{a} di Salerno, I-84081 Baronissi (SA), Italy}
\author{G. Ghiringhelli}
\affiliation{CNR-INFM Coherentia, Complesso Universitario di Monte
Sant'Angelo, Via Cintia, I-80125 Napoli, Italy}
\affiliation{Dipartimento di Fisica, Politecnico di Milano, piazza
Leonardo da Vinci 32, I-20133 Milano, Italy}

\date{\today}

\begin{abstract}

The magnetic and electronic modifications induced at the
interfaces in (SrMnO$_{3}$)$_{n}$/(LaMnO$_{3}$)$_{2n}$
superlattices have been investigated by linear and circular
magnetic dichroism in the Mn L$_{2,3}$ x-ray absorption spectra.
Together with theoretical calculations, our data demonstrate that
the charge redistribution across interfaces favors in-plane
ferromagnetic (FM) order and $e_{g}(x^{2}-y^{2})$ orbital
occupation, in agreement with the average strain. Far from
interfaces, inside LaMnO$_3$, electron localization and local
strain favor antiferromagnetism (AFM) and $e_{g}(3z^{2}-r^{2})$
orbital occupation. For $n=1$ the high density of interfacial
planes ultimately leads to dominant FM order forcing the residual
AFM phase to be in-plane too, while for $n \geq 5$ the FM layers
are separated by AFM regions having out-of-plane spin orientation.
\end{abstract}


\maketitle

Interfaces between different transition metal oxides (TMO) have
been widely demonstrated to be sources of interesting and
unexpected electronic and magnetic properties. Metallic
conductivity arises, for example, at the interface between two
insulators, such as LaAlO$_{3}$/SrTiO$_{3}$\cite{1} and
LaTiO$_{3}$/SrTiO$_{3}$\cite{2}, while ferromagnetism (FM) occurs
at the interface between the antiferromagnet (AFM) CaMnO$_{3}$ and
the paramagnet CaRuO$_{3}$ \cite{3}. In this context, strain
driven spin-orbital coupled states arising in manganites make the
interfaces between these compounds very interesting for
engineering unique collective states. As a matter of fact, a
certain amount of theoretical and experimental studies on
superlattices (SLs) composed by the two AFM insulators,
SrMnO$_{3}$ (SMO) and LaMnO$_{3}$ (LMO), appeared in literature
during the last years \cite{4,5,6,7, 10, Lin, 9, 13, 14}. The
ordered sequence of the atomic layers in the digital SMO/LMO SLs
\cite{4,5}, together with the electronic reconstruction arising
from the interfacial Mn$^{3+}$/Mn$^{4+}$ mixed valence, give rise
to peculiar transport, magnetic and orbital properties, when
different layering and strain conditions occur. In the particular
case of (SMO)$_{n}$/(LMO)$_{2n}$ the La:Sr ratio is 2:1, in
analogy with the optimal composition of
La$_{2/3}$Sr$_{1/3}$MnO$_{3}$ (LSMO). In such a case, the
metal-insulator transition (MIT) and the  magnetic properties
depend on the thickness of the constituent blocks \cite{4,5,6,7},
although in a non trivial way. Indeed, saturation magnetization
does not linearly decrease with $n$ \cite{4} and both fast and
viscous spin populations are present, the latter associated to
FM/AFM pinning \cite{10}. Therefore, the development of the FM
metallic phase at the interfaces is well established and the
coexistance of the FM and AFM phases was inferred. However, the
knowledge of the mutual dependence of the AFM and FM phases with
$n$ is still uncertain, but it could open further perspectives in
the control of the low dimensional magnetic properties, thus in
the engineering of the TMO magnetic heterostructures. In addition,
as the role of interfacial Mn $e_{g}$ electrons is known to be
important, the influence of strain and reduced dimensionality on
the transport properties requires some attention \cite{6, 7, Lin,
9}. In the case of LMO/SMO SLs the two materials widely change
their behavior with respect to bulk. So, while in bulk the $e_{g}$
levels of SMO are empty and LMO presents in-plane
$(x^{2}-r^{2})/(y^{2}-r^{2})$ orbital order, at SMO/LMO interfaces
strain and electronic reconstruction are expected to modify
orbital population and ordering in
both components \cite{13,14}.\\
We have determined the orbital and magnetic properties of
(SMO)$_{n}$/(LMO)$_{2n}$ SLs with $n=1$, 5, 8, by measuring x-ray
\emph{linear} dichroism (XLD) and \emph{magnetic circular}
dichroism (XMCD) in x-ray absorption spectra (XAS) at the Mn
L$_{2,3}$ edge. Thanks to our experimental techniques, we have
been able to follow the evolution with $n$ of the FM and AFM
phases along with the preferred $e_{g}$ level orbital occupation.
We have found that the AFM spin direction is aligned with the FM
easy-axis direction for $n=1$, while is perpendicular to it for
n=5 and 8, and that the preferential orbital occupation is of the
$e_{g}(x^{2} - y^{2})$-type for n=1 and of the $e_{g}(3z^{2} -
r^{2})$-type for $n=5$ and 8. We discuss these experimental
results in terms of the role played in the system by the charge
carrier delocalization and the epitaxial strain.\\
The measurements have been performed at the beam line ID08 of the
European Synchrotron Radiation Facility (Grenoble, France), which
is based on AppleII undulator source (allowing a full control of
linear and circular polarization) and a high scanning speed
spherical grating monochromator (Dragon type). The absorption
signal has been measured in total electron yield. A magnetic field
parallel to the incident beam was provided by a superconducting
electromagnet. The XMCD signal is proportional to the atomic
magnetic moments in ferromagnetically ordered samples
\cite{Stohr}. On the contrary XLD can be due to an orbital or
magnetic uniaxial anisotropy (or both), and the magnetic part
(XMLD, x-ray \textit{magnetic linear} dichroism) is sensitive to
both FM and AFM ordering \cite{Alders}. To disentangle the
magnetic from the orbital part of XLD we have performed
measurements at different temperatures $T$: above the magnetic
ordering temperature only the orbital contribution survives, so in
our samples XLD at high $T$ (room $T$ for $n=1$ and 200K for
$n=5$, 8) have orbital character only. This orbital XLD is due to
the uneven population of the two $e_g$ orbitals (($3z^{2}-r^{2})$
and $(x^{2}-y^{2})$) at Mn$^{3+}$ sites. To obtain the purely
magnetic XMLD signal at low $T$ (10K), we have subtracted the
XLD$_{\textrm{HighT}}$ from the XLD$_{\textrm{LowT}}$, under the
hypothesis that the orbital XLD does not depend on $T$. Moreover,
using a strong (1 tesla) magnetic field parallel to the incident
beam, the FM moments can be fully aligned so that they do not
contribute to XMLD, and the pure AFM contribution to XMLD can be
measured \cite{ArutaXMLD}.

\begin{figure}
\includegraphics[width=8.5 cm]{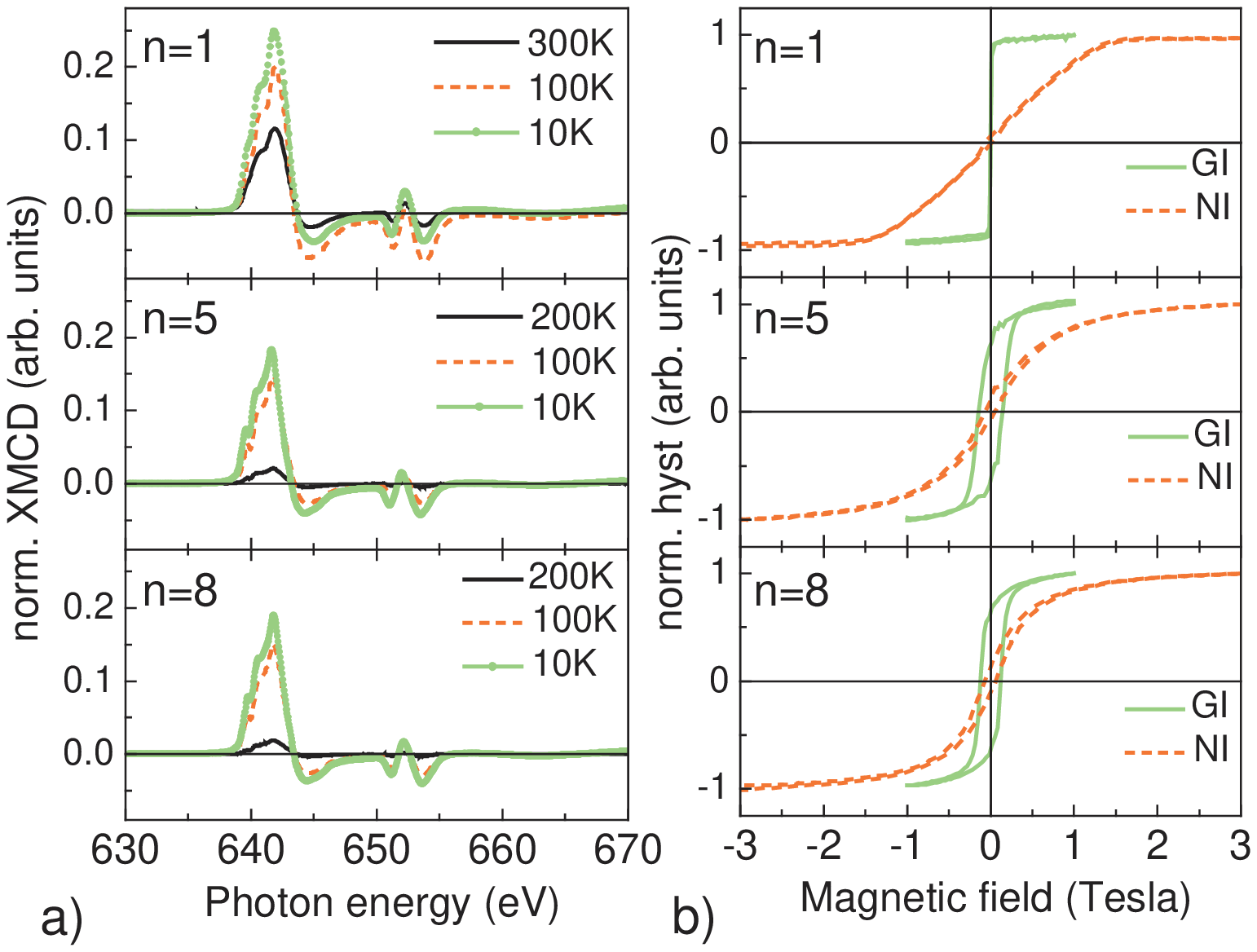}
\caption{(Color online) (a) XMCD results at grazing incidence with
an applied magnetic field of 1 T and at different temperatures for
SLs with $n = 1$, 5, 8. The XMCD results are normalized to the sum
of the XAS $L_{3}$ peak height signals (b) Hysteresis loops at 10
K obtained by the maximum peak intensity of the XMCD (about 642
eV) as a function of the applied magnetic field. The curves are
normalized to unity for a better comparison of the coercive
fields.}
\end{figure}

The investigated samples were all grown on SrTiO$_{3}$ (STO)
substrates and with a total thickness of 200\AA\cite{4}. The
crystallographic structures of the SMO and LMO constituent blocks
are different, between them and with respect to the STO substrate.
As a consequence, the (SMO)$_{n}$/(LMO)$_{2n}$ SLs epitaxially
grown on STO experience a modulated strain, with the SMO and LMO
layers being in-plane compressed and tensile extended,
respectively. In particular, bulk LMO is an A-type AFM \cite{21}
and the room-temperature crystal structure belongs to the
orthorhombic space group \textsl{Pbnm} with lattice constants
$a=5.537$ $\AA$, $b = 5.747$ $\AA$ and $c = 7.693$ $\AA$
\cite{20}. Bulk SMO is a G-type AFM \cite{22} and the cubic
lattice cell belongs to the \textsl{Pm3m} \cite{23} space group
with lattice parameter $a=3.805$ \AA. Both LMO and SMO films try
to release the stress energy induced by the large lattice mismatch
between the film and subtrate and partially relax their epitaxial
strain. Indeed, 200\AA thick films of LMO and SMO on STO substrate
have shown $c$-axis values about $3.93 \AA$ and $3.78 \AA$,
respectively. Very strained interfaces are therefore formed in the
SLs, where the FM phase nucleates and starts to propagate far from
the interfaces in the AFM constituent blocks. Such interfacial FM
phase is detected in all investigated samples by the XMCD
measurements of Fig.1, which is added to some FM content of the
LMO block already discussed in literature \cite{4, 5}.

\begin{figure}
\includegraphics[width=6.5 cm]{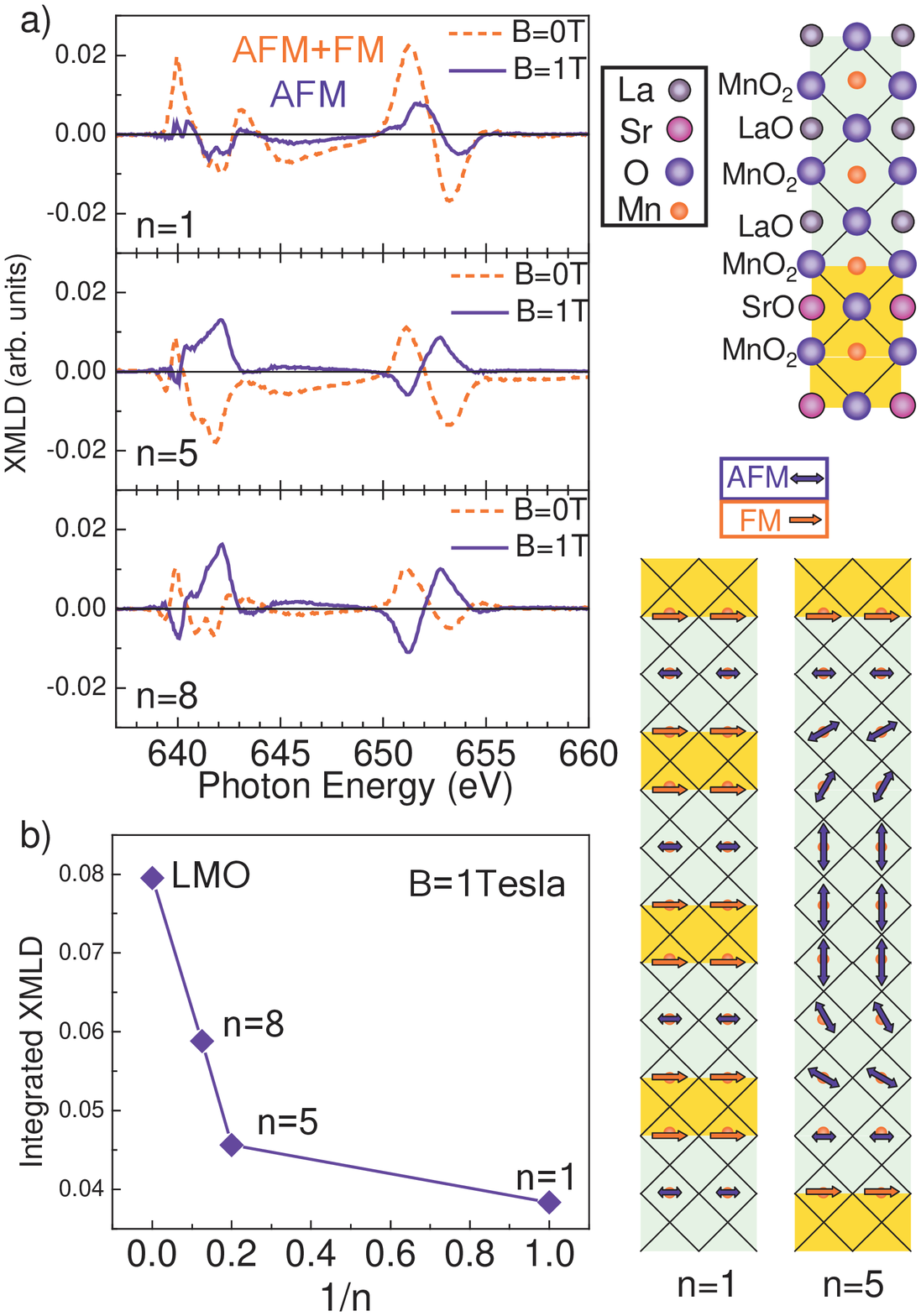}
\caption{(Color online)(a) Difference between the XLD spectra
taken at 10K and 300K ($n=1$) or 200K ($n=5$, 8), with B = 0 T and
B = 1 T. All spectra are normalized to the sum of the XAS $L_{3}$
peak height signals. (b) Integral of the curves at B = 1 T with
respect to $1/n$. On the right, the schematic drawing of the local
spin orientation in the superlattices is reported. The length of
the arrows is roughly proportional to the magnetic (AF or AFM)
content of the $MnO_{2}$ layers.}
\end{figure}

In Fig. 1b the hysteresis loops in grazing (GI) and normal
incidence (NI) configurations are also reported. It can be
observed that FM easy-axis is mostly oriented in the $ab$-plane.
This FM anisotropy resembles that one induced by the
magneto-crystalline anisotropy in the LSMO films under tensile
strain \cite{Nath}. The most interesting result is that the
anisotropy ratio is larger in the SL with $n=1$, where the density
of FM interfaces is highest, and decreases with $n$. Large
hysteresis cycles as in our $n=5$ and $n=8$, already reported in
literature for the $n$ = 2, 3 and 5 \cite{10}, were explained in
terms of competing AFM/FM interactions with magnetic pinning,
frustration and canted order. Somehow more surprising is that in
all samples we find an important AFM contribution to XLD as shown
in Fig. 2. As explained above, for $B=0$ T both FM and AFM phases
contribute to XMLD whereas for $B=1$ T the signal is coming only
from the AFM phase. For $n=5$, 8, and for pure LMO (not shown)
XMLD(0 T) and XMLD(1 T) have similar shape but opposite sign,
while for $n=1$ they have the same sign. This means \cite{Fujii,
Freeman} that in the former cases the FM and AFM phases have
magnetic moments oriented in (roughly) orthogonal directions, but
are approximately parallel for $n=1$. Therefore, as the FM
easy-axis is mostly in-plane in all our samples (as confirmed by
XMCD measurements of Fig. 1), we can conclude that the AFM
easy-axis is in-plane for $n=1$ and out-of-plane in the other SLs.
Out-of-plane local spin AFM direction has been also observed on
LMO single films (not shown). We notice from Fig. 2 that XMLD(1T)
is weaker for $n=1$ and it increases with $n$, indicating that the
fraction of AFM phase increases when reducing the interface
density. In Fig. 2b we summarize these information by plotting the
integral of the XMLD spectra at B = 1 T. Observing that for  $n=1$
the single SrO layers are sandwiched between two interfaces,
namely the interfacial MnO$_{2}$ planes, the intrinsic properties
of SMO block are not recovered. On the contrary, for bigger values
of $n$ an AFM contribution from the SMO layer could survive, as it
has been theoretically predicted by Nanda and Satpathy \cite{7}
together with a contribution from the LMO block. Also in several
experimental works \cite{5, May, Smadici} the AFM region has been
located in the SMO blocks too. The schematic drawing on the right
side of Fig. 2 depicts that, for very thin constituent blocks like
$n=1$, the AFM local spin direction is pinned to a mainly in-plane
orientation by the interfacial FM anisotropy. The AFM content is
responsible for the reduced Bohr magnetons number in SL $n=1$,
about 3.0 $\mu_{B}$/Mn \cite{4} compared with the optimal 3.7
$\mu_{B}$/Mn value of the LSMO films \cite{Orgiani}. When
increasing the constituent blocks thickness the AFM phase might
remain oriented in-plane only very close to FM interface, while it
keeps the out-of-plane anisotropy far from the interfaces.\\

\begin{figure}
\includegraphics[width=6.5 cm]{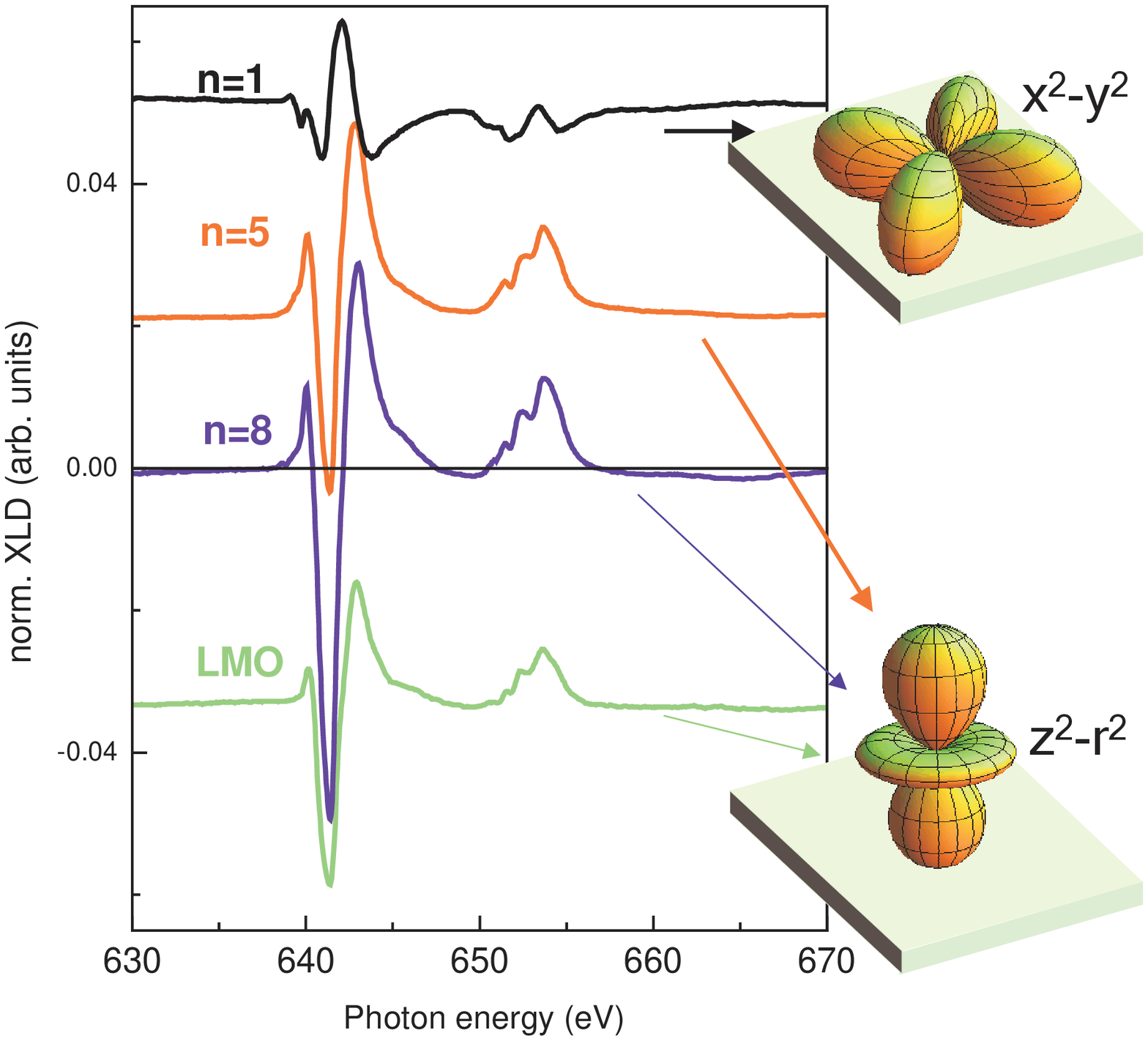}
\caption{(Color online) XLD spectra at 300K for $n=1$ SL and 200K
for $n=5$, 8 SLs and the LMO film. The spectra are reported as the
difference of the XAS measurements with Vertical (V) and
Horizontal (H) polarizations in grazing incidence configuration
and normalized to the sum of the XAS $L_{3}$ peak height signals.}
\end{figure}

A further insight in the SLs properties comes from the XLD spectra
above the magnetic ordering temperature. The XLD spectra of Fig. 3
show that for $n=5$ and $n=8$ the preferential orbital occupation
is the same of pure LMO on STO: out-of-plane $e_g$ orbitals
($3z^{2}-r^{2}$ or $(y^{2}-z^{2})$/$(z^{2}-x^{2})$) are
preferentially occupied, in analogy to what has been found for
LSMO films under compressive strain \cite{15, TebanoPRB}.
Moreover, we observe that the XLD amplitude grows with increasing
$n$, i.e., with decreasing interface density. On the contrary, for
the metallic $n=1$ SL the XLD signal, although weaker, is clearly
reversed in sign with respect to the other cases, indicating an
in-plane preferential orbital occupation, possibly with
$e_{g}(x^{2}-y^{2})$ symmetry \cite{15, EPLAruta}.

\begin{figure}
\includegraphics[width=6.5 cm]{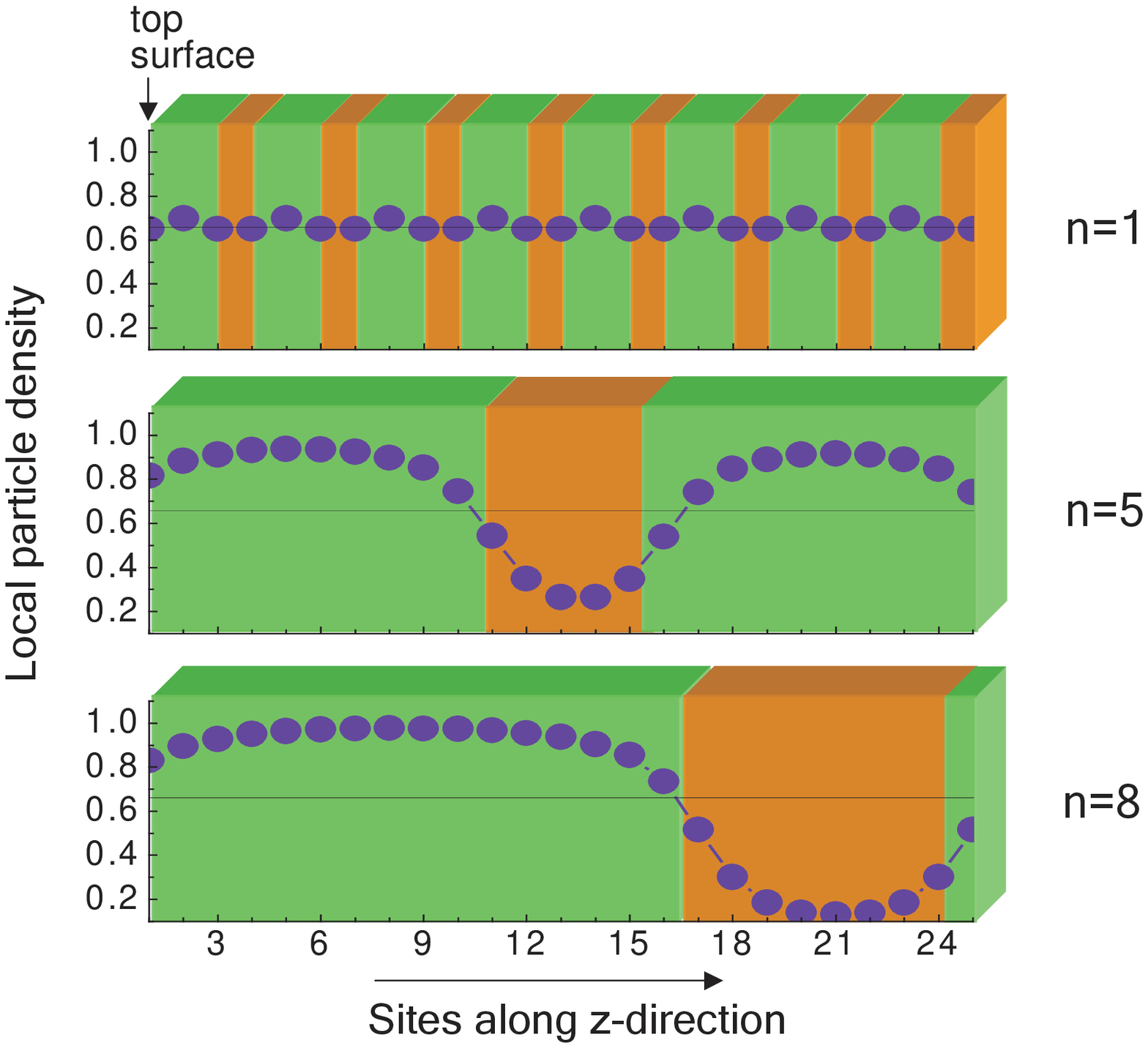}
\caption{(Color online) Schematic drawing of the top part of SLs
measured by total electron yield, together with the spatial charge
density (blue dots) along the samples, calculated as described in
the text. Green and orange zones represent LMO and SMO constituent
blocks, respectively.}
\end{figure}

These results can be understood starting from the following
observations. Firstly all superlattices are coherently strained:
all of them are forced to the in-plane lattice parameter $a$ of
the STO substrate and to an average out-of-plane parameter $c
\simeq 3.87$ \AA \cite{4}, thus giving $c/a<1$ in average. As a
consequence, the LMO blocks are subjected to compressive strain
($-2.2\%$) and the SMO blocks to tensile strain ($+2.6\%$).
Furthermore, the orbital contribution to the XLD for $n=5$ and
$n=8$ is mainly given by the LMO layer since in those SLs the Mn
sites in SMO layers are essentially $3d^3$, a configuration that
is spherical and cannot contribute significantly to XLD.
Calculations of the spatial charge density for the three
superlattices (Fig. 4) indicate that the $e_{g}$ levels of
Mn$^{4+}$ in SMO are generally not occupied apart from a narrow
region at the interface. On the contrary, the compressive strain
on LMO block of the SLs, where Mn is $3d^4$, stabilizes the
$(3z^{2} - r^{2})$ orbitals, leading to a dichroic signal similar
to that of LMO alone. The results of Fig.4 are based on density
calculations made within a self-consistent Hartree approach for
the SL configurations (see also \cite{Lin, 6}), and on the results
of Ref.\cite{9} . The previous arguments can also explain the
stronger XLD signal in the $n=8$ SL with respect to the $n=5$
case. In the latter case, in fact, the $e_g$ electron density
distribution, as shown in Ref.\cite{9}, shows a small reduction in
the LMO region and an increase in the SMO (see again Fig.4). The
former reduces the $e_{g}(3z^{2} - r^{2})$ signal while the
latter, subjected to tensile strain, contributes with a very small
$e_{g}(x^{2}-y^{2})$ component. For $n=1$ the situations is
completely different. The contributions from LMO and SMO blocks
cannot any longer be distinguished, the $e_g$ electron density
distribution becomes almost constant (see Fig. 4) and equal to the
average density. The strain does not act in opposite way on LMO
and SMO, but the system responds as a whole to the average strain
that is slightly tensile, and the $e_{g}(x^{2}-y^{2})$
orbitals get stabilized as in LSMO grown on STO\cite{15,EPLAruta} or in (LMO)$_{1}$/(SMO)$_{1}$ SLs \cite{14}.\\

In conclusion, our study demonstrate that when the charge transfer
through the interfaces delocalize the e$_{g}$ electrons into the
entire SL, the $n=1$  SL behaves as an homogeneous system. On the
contrary, when the interface density is smaller ($n=5$, 8) and the
distance between interfaces is higher than the Thomas-Fermi length
\cite{Lin, 5}, the role played by the strain applied to each
constituent block becomes fundamental. As a consequence the
electronic localization is accompanied by the preferential
out-of-plane orbital occupation. Moreover the in-plane easy-axis
of the double-exchange FM spin orientation, with the pinned AFM
spin orientation, further confirms that the uniform electronic
distribution in the $n=1$ SL causes properties dominated by the
average strain effects as a whole, thus favoring the in-plane
orbital occupation. However, when the thickness of the constituent
blocks increases the interfacial FM phase does not extent in the
whole superlattices and the AFM spin direction starts to rotate
towards the out-of-plane direction.\\

C. Aruta acknowledges the ESF activity Thin Films for Novel Oxide
Devices (THIOX), within the Exchange Grants program.

\end{document}